# CYBER SECURITY INSIGHTS INTO SELF-PROCLAIMED VIRTUAL WORLD HACKERS

Nicholas Patterson, Michael Hobbs, Frank Jiang
and Lei Pan

School of Information Technology, Deakin University, Geelong, Australia

## ABSTRACT

*Virtual worlds have become highly popular in recent years with reports of over a billion people accessing these environments and the virtual goods market growing to near 50 billion US dollars. An undesirable outcome to this popularity and market value is thriving criminal activity in these worlds. The most profitable cyber security problem in virtual worlds is named Virtual Property Theft. The aim of this study is to use an online survey to gain insight into how hackers (n=100) in these synthetic worlds conduct their criminal activity. This survey is the first to report an insight into the criminal mind of hackers (virtual thieves). Results showed a clear-cut profile of a virtual property thief, they appear to be mainly aged 20-24 years of age, live in the United States of America, while using virtual worlds for 5-7 hours a day. These and the other key results of this study will provide a pathway for designing an effective anti-theft framework capable of abolishing this cyber security issue.*

## 1. INTRODUCTION

The problem related to cyber security is becoming increasingly challenging due to escalating security attacks on networks [1] as evident by recent attacks by the hacking group 'Anonymous' [2] and the director of the Federal Bureau of Investigation stating that cybercrime will eclipse terrorism [3]. The issues involving cyber security brings with it many dilemmas; does cybercrime denote the emergence of a new form of crime or criminality? The major issue with cyber security offences like hacking is whether crime follows opportunity, specifically the criminal exploitation of internet technologies such as online payments, auctions, gaming, social networking sites [4] and in the case of this paper virtual worlds.

There is an ever-increasing popularity with second generation internet-based services which encourage online collaboration and sharing among users, for example Web 2.0 and virtual communities [4] such as the virtual world Second Life [5]. Virtual Worlds Environments (VWEs) are essentially computer based simulated environments whereby users can create an avatar (digital character) to represent themselves and then enter into the 'virtual world' and interact with other users (or their avatars) in real time, they can participate in games, discussions and professions, as well as an economy based on the trading of virtual world currency and goods. In this paper we define VWEs as an online persistent social space, which aims to simulate the real world as much as possible to create an effect of immersion.

As mentioned users engage in an economy based on virtual property items and virtual currency; these items play a key role in improving the ability or visual appearance of avatars, due to this effect they can take a combination of time and effort to gather. This requirement of time and effort has brought forth many users who are more than willing to pay 'real money' to buy virtual property so that they can upgrade their avatars [6] and increase their status in the virtual world [4]. The fact that users are purchasing these items has led to detrimental effects on VWEs such as increases in online cheating, theft, robbery and much more [6]. The most common cyber security issue that is occurring in VWEs and the highlight of this paper is the offence of Virtual Property Theft (VPT); this entails the online digital theft of another user's virtual property items from within a VWE. This theft can originate from such acts as hacking, deception or trickery and sometimes even robbery. The population of VWEs has increased dramatically over the past few years, Watters [7] announced that registered users of virtual worlds reached one billion worldwide. With regards to the market value of virtual property sales, virtual world's expert Marcus Eikenberry estimated it to be between 10 and 50 billion US dollars [8]. Many users of VWEs have invested large amounts of real money to gain virtual property in their virtual accounts and VPT is not just costing the users in terms of loss of virtual property but also in terms of real money. Taken together, the growth in VWE users, growing virtual property value and increasing cases of theft, there is a need for a deeper understanding of VPT and the development of adequate intervention.

To provide evidence that the crime of VPT is making its mark on both the virtual world and the 'real world' we present some relevant cases below which not only resulted in financial problems for the stakeholders involved (developers, publishers and users), but even resulted in physical assaults being perpetrated against the users. In a recent case of robbery revolving around VWEs was in the Netherlands [9], a teen was sentenced for stealing virtual property items. This case involved a teenage boy beating (kicking the head and body and trying to strangle) another teenager in the victim's own room and then proceeding to steal his virtual property items. The theft occurred after the victim had been beaten, when the attacker threatened the victim with death if he did not agree to log into his virtual world account in order to transfer all his virtual currency and virtual property items to his assailants account. The Netherlands judge in this case stated that "Goods don't have to be material for the law to consider them stolen" and that "Stealing virtual goods is a crime" [9]. The judge sentenced the offender to 160 hours of unpaid work or jail time of 80 days [9].

Another case of virtual offences resulting in criminal charges was in 2008, when a 43-year-old woman hacked into the computer of a man she married from within a VWE, and then proceeded to erase his avatar that he had spent a lot of time and effort creating [10]. She did this because their online relationship and marriage had ended, and she was not happy about it. The offender in this case faces the charge of using her victim's username and password to illegally access the victim's computer, the charge carries a maximum of five years in prison or a fine of approximately 4,165 US dollars [10].

Surveying of self-confessed offenders as well as reporting of cybercrimes has been examined in the past [11] [12-14]; this work primarily has been focused on offenders of traditional or physical act crimes such as youth violence, burglary, assault and sex offences. To our knowledge, there are no self-reporting offender surveys of hackers or cyber criminals; the closest was a study [15] which examined the differences between individuals self-reporting computer-related deviant behaviour and those reporting no computer-related deviant behaviours. This related work will be covered in the discussion section.

## 2. METHODOLOGY

In order to address the problem of VPT in VWEs, this study will provide a mixture of quantitative and qualitative methods to gather and analyse data obtained from a survey completed by VPT self-proclaimed offenders (n=100).

A survey consisting of multiple sections was designed for this research study to determine individual factors on how VPT occurs from an offender's point of view. Participants were sought from a variety of online virtual world community groups (forums/message boards) as this provided a way to reach a large audience of virtual world users and potentially virtual property thieves. This is not an uncommon approach with the following study collecting cyber threat intelligence from hacker forums [16] and a cross-community analysis of 12 hacking forums [17]. The survey was conducted in two stages; an initial pilot study was run to validate the questions, which was then followed by the live survey. The survey link was sent out to many popular online virtual world community groups around the world. The survey was completely anonymous and survey respondents were asked to complete the survey voluntarily with no incentives or rewards. Ethics approval was obtained under Deakin University, ethics ID: 2011-166.

## 2.1 SURVEY

The process used in this study involved receiving input from a number of respondents (n=100) and was focused towards virtual property thief offenders, to gain an understanding and gather views on their experiences regarding the act of conducting VPT. The delivery method for the survey was the internet; this allowed the survey to reach a greater audience spanning many different countries other than Australia.

The survey instrument was 7 pages of multiple choice questions that were divided into sections that gather information on virtual world profile analysis, thief profiling, theft practices, recovery and detection, and security analysis. We provided them with a brief description indicating what virtual property and VPT was. This description is provided below "Virtual property theft is the act of breaking into virtual world user's account and stealing virtual property goods/items and virtual currency (gold); often to be sold on black markets for real money. Essentially the theft and sale of stolen digital objects for real world money."

An ethics plain language statement was presented at the beginning of the survey for participants to view before commencing.

## 3. SURVEY RESULTS AND ANALYSIS

This section provides the results and analysis of the survey from the respondents. A total sample size of one hundred (n=100) respondents were used in this study. Results are presented in the form of graphs and tables. Graphed and tabulated results are expressed as a frequency digit of the number of respondents. In certain graphs where an option is shown as 'other' indicates that this result is a grouping of a multiple choices, a collection of results that represented a very small minority from the total response group.

### 3.1 VIRTUAL WORLD PROFILE ANALYSIS

This section presents analysis of the responses from the first set of questions related to the demographics of offenders and their choices in relation to which VWEs they use and how often they use them. Questions covered in this section include: the age of respondents; the region of the world respondents are from and how long they have been using VWEs for (in terms of months/years and daily usage).

#### 3.1.1 Age of offenders

Figure 1 shows the results of the age of respondents who participated in this survey. Each response is listed as a frequency digit that represents the age of the respondents.

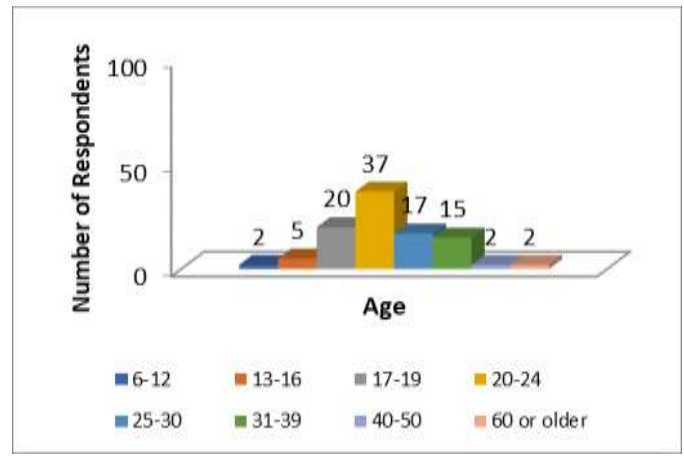

Figure 1.  The age of respondents who are self proclaimed offenders. (n=100).

It was an important aspect of this study to discover what the average age of offenders and determine if it matches with the average age of video game players or is it totally different. To gain an understanding of video game players, the results of a 2016 Digital Australia Report [20] were used, the results of this study relate to Australian video game players only but still provides us with a sample view into the population.  It shows that the average age of an Australian video game player was 33 years and 7 out of 10 are male. The age of VPT offenders presented in the graph show a high response rate between the ages 20-24 being the most dominant at 37 responses, followed by 17-19 years of age at 20 responses As the age result from our survey does not link closely with this study of [20], this suggests that virtual property thieves may not fit the profile of an average video game user.

### 3.1.2  Offenders region of the world

Figure 2 shows the geographical location of the respondents of the survey. The response from each respondent is listed as a frequency digit that represents the region of the world they are from.

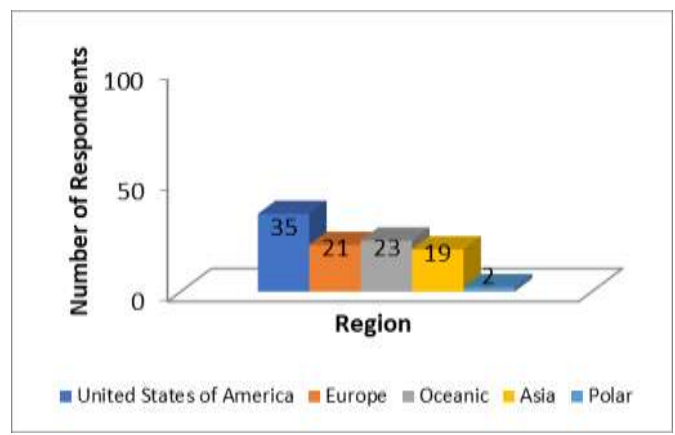

Figure 2. Regions of the world respondents who are self proclaimed offenders are from. (n=100).

The region of the world in which VPT offenders originate from is presented in the graph (Figure 2); shows that out of the 100 respondents, the dominant response was United States of America (USA) at 35 responses, followed by Oceanic regions at 23 and Europe at 21. These results indicate that respondents who originate from USA may be considered the most likely offenders of VPT by a vast amount compared to other regions presented in the results. There is a 12 point margin between USA and the next highest region being Oceanic. From these results there may be increased focus on targeted analysis of users from USA, by means of greater analysis of authentication and activity data from this region.

## 3.2 THEFT PRACTICES

The next section of questions is related to theft practices and activities employed by offenders. Questions included: what the reason was, when and why offenders chose specific times of the day or night to conduct VPT, the amount of virtual property and virtual currency stolen, which items these offenders stole first after breaking in and if all the virtual property was stolen in one or multiple break ins.

### 3.2.1 Age of offenders

There was a wide variation of times chosen to conduct VPT with a few popular times indicated in the graph. The most dominant choice from respondents in response to the reason for choosing a particular time to conduct VPT was 6PM-10PM with 19 responses, followed by 10AM-12PM with 15 and 3PM-6PM with 14. These results can indicate two factors, the first is that thieves are not distinctly choosing any particular time to conduct VPT, which may indicate it's random or just whenever the offender has free time to perform the offence. The second indication may be that the chosen times indicate that offenders of VPT are choosing what could be considered 'peak times' for VWE usage; therefore, it may also indicate that offenders are not attempting trying to avoid the potential for the owners of the account (being broken into) to be logged in and conducting their normal activities.

### 3.2.2 Why time for theft was chosen?

The aim here was to discover if there was in fact any reasoning towards choosing the time offenders conduct VPT or was it in fact totally random as suggested. The reason for choosing this time is listed as a frequency digit that represents the choice of respondents. The most dominant choice presented by respondents with relation to why they chose particular times to conduct VPT was that it's 'the only time I have free' with 51 responses. This demonstrates that offenders are not concerned about interacting with the owners of these accounts they break into, and will pursue the act of VPT regardless. It can also be shown that offenders may in fact be no different than normal users and will use VWEs at the same times as everyone else, just with a different purpose, being that of theft of virtual property items.

### 3.2.3 Amount of virtual property stolen

We directed a question to respondents on how many virtual property items they had stolen over the course of their VPT activities. The amount of virtual property stolen is listed as a frequency digit that represents the choice of respondents. The goal of this question was to find out how many virtual property items are stolen in general and with comparison to virtual currency. The most dominant choice from respondents was '1000s of items' at 26 responses. The lowest response was from offenders who indicated they had stolen '5 items or less' at 2; it could be seen that these may be first time offenders. These results indicate a high rate of item theft by offenders, they are simply not going in and stealing a couple of virtual property items, but appears as though they are stealing vast quantities of what may be valuable virtual property items.

### 3.2.4 Amount of virtual currency stolen

We wanted to determine how much virtual currency had respondents stolen over the course or lifetime of their virtual world activities, this is not how much was stolen in one instance. When it comes to virtual currency, it is the most important item that can effect revenue and there are 2 types: one is a paid virtual currency (PVC) obtained by billing and the other is a free virtual currency (FVC) which is freely distributed through accomplishment of tasks [21].

The amount of virtual currency is listed as a frequency digit that represents the choice of respondents. The goal of this question was to show a relationship with responses from the previous question which asked respondents *what the amount of virtual property they had stolen was*. This way a view can be given to see what the comparison is between virtual property items stolen and virtual currency stolen. There was a wide variation of responses for this question, which indicates many varying amounts of virtual currency stolen. It is common in VWEs to start new users off with a small amount of currency that is just enough to get them started in the virtual world. This provides a motivation to 'play' and to further progress with their in-world character. This is why there is such a large market for the sale of virtual currency within online markets, purchasable with real world money; it essentially allows player to skip or avoid the spending of time and effort to collate virtual currency.

The amount of virtual currency stolen is substantially more than virtual property items stolen. There was a high grouping of respondents being 19 who had stolen 'over 30,000' units of virtual currency, followed by 15 with a substantial increase of 'over 1,000,000' virtual currency units.

These results indicate that offenders vastly prioritize stealing virtual currency over virtual property. This difference is due to a to a number of possible reasons such as virtual currency generally being a more abundant commodity. This is because virtual currency is fundamental in all VWEs and is used for buying and selling virtual property items and used as a reward from in-world activities such as quests or missions.

## 3.3 Sale of stolen virtual property

This section analyses the responses from the questions related how respondents sold stolen virtual property items and how much they have made in terms of profit. The questions posed in this section ask offenders if they have they sold stolen virtual property items on primary markets or secondary markets and how much money have they made through these sales.

### 3.3.1 Stolen virtual property sold on primary markets?

This section relates to results from the question that asked respondents, have you sold stolen virtual property items on primary markets (which are most often used for legitimate sales, due to much more stringent security procedures). Their ability to sell on these primary markets is listed as a frequency digit that represents the choice of respondents. In this question and the following question, the aim was to determine what markets respondents were selling stolen virtual property items on and as a result answer a number of questions such as; are offenders able to sell stolen items on legitimate markets or more so just on secondary markets (where stolen items are much easier to sell due to less stringent security procedures).

There were 51 respondents which claimed 'yes' they have sold stolen virtual property on primary markets owned/operated by Habbo Hotel, Second Life, Gaia Online, Sony Station Exchange and Entropia Universe. These primary markets are more stringent with their security and identification measures, often involving the buyer/seller registering their license or credit card with the system and using a legitimate account linked to the VWE or utilizing a form of escrow system (where the money is held by a middle man until the transfer is confirmed by both parties). It seems offenders of VPT can and are using these markets for selling stolen virtual property items, demonstrating that they are finding avenues to avoid this extra security or identity checks.

### 3.3.2  Stolen Virtual Property Sold on Secondary Markets?

Here we asked respondents 'have you sold stolen virtual property items on secondary markets' which are most often used by buyers/sellers of virtual property items not requiring a need for stringent security and identification procedures. The reason for this is that the primary markets due to providing better security and identity check measures, take a significant percentage of the sale price, these secondary markets often either don't take a cut of the sale price or it is very low. The ability to sell on these secondary markets is listed as a frequency digit that represents the choice of respondents. This question is a follow on from the previous question; however here the aim was to discover how many offenders are selling stolen virtual property items on secondary markets or what are often referred to as black markets in the RMT (Real Money Trading) market. These markets include some of the following: PlayerAuctions, IGE, YahooAuctions and in the past eBay (before they banned the sale of virtual property in 2007). In these markets it is much easier for an offender to sell stolen items due to more lax security and identification measures or the ability to bypass those measures.

There are 76 offenders who claim to have sold or are selling stolen virtual property on these secondary markets, as little as 24 say they have not. These results were expected as mentioned previously in the introduction to this; it is much easier for offenders to sell stolen items without being identified or discovering that the items being sold are indeed stolen as no link to the actual VWE is active.

### 3.3.3  Real Money Trading of Stolen Virtual Property

Figure 3 shows the results from the question that asked respondents, how much 'real world' money (USD$) have you made from selling stolen virtual property items. The amount of money made from sales is listed as a frequency digit that represents the choice of respondents. This aim of this question was to determine how much real world money has been made by offenders of VPT. This helps to determine factors such as the extent of the RMT market and if offenders of VPT are selling stolen items or using them for themselves or given to associates.

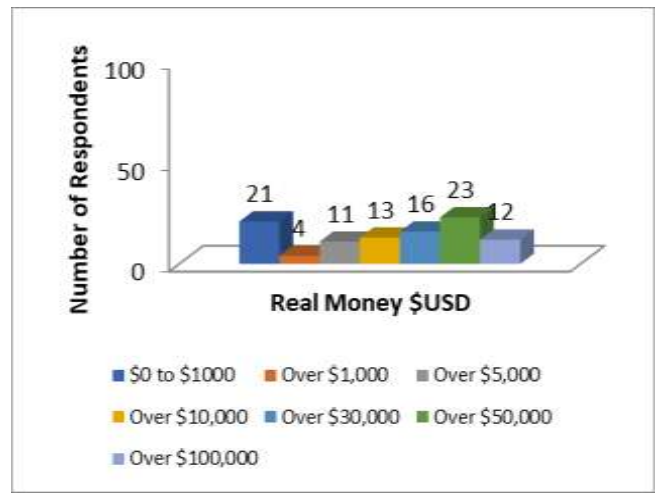

Figure 3. How much real world money is made by offenders (n=100) from the sale of stolen virtual property items.

As shown in Figure 3 the RMT (real money trading) market is quite large, the most dominant response was 23 respondent's stated they have made over 50,000 US dollars. There was a grouping of 21 respondents who claim they have made under 0 to 1000 US dollars, this grouping are believed to be first time offenders, first time sellers or newly started offenders. The total amount of money made from all respondents was 5,221,000 US dollars (this was calculated by adding up the total amount values given by each response) or in rudimentary terms upwards of 5 million US dollars.

There have been certain cases…for example in the largest online sale of virtual property, a virtual nightclub named 'Club Neverdie' in the VWE Entropia Universe sold for 635,000 US dollars [22]. Another example demonstrated virtual property worth in the real world, where a high school student and his mother made 35,000 US dollars by creating, farming and selling virtual weapons and animal skins within VWE Entropia Universe in 2006 [22, 23].

## 3.4  Recovery and detection

This section relates to recovery and detection; to be more specific when a virtual property item is stolen occasionally it can be recovered by specialist staff employed by the virtual world company running the VWE, this can be a difficult task as virtual property items and currency can change hands very fast and a victim may not realise they have had items stolen until days after. Detection however relates to the ability for virtual world operators (often referred to as administrators or game masters) or the VWE software itself to detect virtual property thieves and intruders. This section analyses responses from the questions related to theft, recovery and detection of VPT. The goal of these questions is to identify the issues, activities and measures taken with relation to recovery and detection that are directly related to VPT among the respondents.

### 3.4.1  The ability to detect theft of virtual property

This section discusses the results from the question that asked respondents, 'do you ever get caught while conducting theft of other user's virtual property items'. The ability to detect theft of virtual property items is listed as a frequency digit that represents the choice of respondents. In this question the aim was to know if the offenders had been detected whilst conducting the theft of virtual property items specifically. This detection would occur when the offenders are trading virtual currency and or virtual property items to an account the offender owns to store stolen virtual property items.

The most dominant response was that 92 offenders are not being detected or caught whilst conducting theft of another user's virtual property items. These results demonstrate offenders for the most part are not being detected when conducting VPT, so they can perform this offence while running a very low risk of being caught and may lead to a high profit as a result.

### 3.4.2 Recovery and return of stolen virtual property

This section highlights the results from the question that asked participants, do the virtual property items you steal ever get returned to the original owner. The rate of return of stolen virtual property items is listed as a frequency digit that represents the choice of participants. In this question the aim was to discover if offenders have found that once they have stolen virtual property items from a victim, if the virtual world operators have detected the theft and then reversed the unauthorised trades and return the virtual property items back to the victim.

The most dominant response being 50 offenders have stated 'no' to virtual property items being returned to the victim after theft. These results demonstrate that stolen items are very rarely returned to the victim, which means when they are stolen they will remain stolen and often resold onto what may be unassuming consumers in virtual property markets.

## 3.5 Security Analysis

This section will analyse the responses from the questions related to ownership of virtual property items and currency from a thief's perspective. Questions will cover: the most useful hacking method offenders use to conduct VPT, self-proclaimed ability to conduct VPT, effectiveness of virtual world operators to stop VPT from an offenders point of view, the effectiveness of security measures available to users to protect themselves against VPT, what kind of computer and internet configuration offenders use while conducting VPT and self-reported ability to break into victims accounts. Each will present results in either text, Figure or a Table, and analysis will follow.

### 3.5.1 Most popular method to gain access

Table I shows the results from the question that asked participants, what they believe is the most successful method for breaking into victims accounts based on their experience. The amount of success of breaking into victim accounts with the chosen tool is listed as a frequency digit that represents the choice of participants. Respondents were allowed to choose more than one response in this question.

Table 1. Listing of most useful hacking methods used to break into virtual world accounts or stealing virtual property items by offenders (n=100).

| Hacking Method / Tool | Frequency of Use |
|---|---|
| Trojan virus on user's computer | 33 |
| Hacked user's email account | 36 |
| Publishing a link on a web forum which downloads a virus to the victim's computer | 28 |
| Hacked the virtual world server | 10 |
| Vulnerabilities in the virtual world software | 37 |
| Guessing or brute forcing a user's password | 37 |
| Social engineering or tricking user's into giving their account info | 49 |
| Virtual property trade scam | 46 |
| Killing and looting avatars in the virtual world | 16 |
| Man in the middle attack | 18 |
| Guess user's security questions | 33 |
| Physically watch a user enter in their login/password | 14 |
| Other (please specify) | 9 |

In this question the aim was to discover from offenders what their most effective hacking method or tool for breaking into victims virtual world accounts was. We used the following cybercrime offences study [24] which provides three separate and complementary views to achieve a comprehensive perspective on cybercrime offences and responses. As seen in Table I the most dominant response in was 'social engineering or tricking user's into giving their account information' at 49 responses, followed by 'virtual property trade scam' with 46. These two options seem to be the most effective for offenders, one being a more technical hacking attack and one seen to be a more social engineering technique. The results shown here demonstrate that offenders are utilizing a wide array of exploits, tools and techniques to break into user's accounts in order to steal virtual property items.

## 3.6 Discussion

In this section we discuss how our results compare and contrast with related studies. This assists in determining things such as the scale of VPT in comparison to other crimes, if cybercrimes and are self-reporting offender surveys are a valid research tool and can finally if results gained from studies of this nature be useful for prevention or cyber security incidents.

The following study [13] focused on analysing the cybercrime which is occurring in Indonesia, this occurred specifically through a literature survey of existing results from various sources. Results show that Indonesia has jumped up by 545 cases of cybercrime in 2011 and growing more in 2012 by 600 cases. The 5 most common cybercrime attacks detailed in the study are malware/spam, data theft, ID theft, phishing and botnets. Comparing the results with our study in Table I we can see that the 4 most common type of hacking methods used to conduct VPT were (from most frequent to least) social engineering, virtual property trade scam, guessing or brute forcing a user's password and discovering vulnerabilities in the platform's software itself. So common linkages between Indonesia's cybercrime results and our results are related to data theft, ID theft and some of our more lesser frequent methods in malware. This study gives a concise view of the cybercrime incidents which are occurring in Indonesia and provides benefit towards seeing where the types of hacking methods used by hackers in VWEs correlate to the cybercrimes commonly reported in Indonesia.

A study to examine the differences between individuals self-reporting computer related deviant behaviours and those reporting no computer related deviant behaviours was conducted by Rogers et al. [15]. In this study they put focus on what they call the 'Big-5' personality characteristics, moral decision making and exploitive manipulative amoral dishonesty characteristics [15]. They discovered in their results that computer deviants scored low on social moral choice and were significantly exploitive and manipulative [15]. Compared with the results of this study, we could determine that offenders of VPT have significant disregard for victims; breaking into accounts during peak periods (when the victim may be active) and stealing not only all the rare or most expensive property but also all the virtual currency the victim owns. They also appear to be extremely deceiving and manipulative with a high majority of offenders using social engineering, trickery or scams to conduct VPT. This study provided some good empirical knowledge into deviant computer behaviour but was more focused from a psychology aspect and not related to actions and behaviours which occur in cybercrimes such as VPT.

Kumar [14] conducted a study which reviews the growth of cybercrimes in India and measures taken by the government of India to combat cybercrime and details the cybercrimes registered as well as persons arrested under the Information Technology Act, 2000. Their results show that there has been a large increase in cybercrime attacks over recent years. However, this increase in attacks has not led to a significant increase in arrests. While there were 4356 cases registered under the IT act, the number of people arrested was under 50% with only 2098 reported arrests. In the study they also discovered that offenders of the cybercrimes reported were mostly from the age group of 18-30 years of age, with the next most common being the 30-45-year-old age group. The top three cybercrimes came under the following categories: forgery, criminal breach and fraud (currency or stamps). To compare with the results in our study, offenders ages who were most prevalent with conducting VPT were 20-24 years of age (n=37). Then also the act of VPT tends to correlate strongly with the act of criminal breach (hacking into a user account and stealing virtual property goods). This study while focused on India and the acts of cybercrime encountered there, provides a good comparative analysis especially when it comes to the age of offenders.

Farrington [25] released a report which summarises what has been learned from the process of self-reporting criminal careers and the causes of offending. They discovered that most knowledge about criminal careers has been solely based on official records of arrests or convictions. Farrington [25] argues that self-reports provide a more accurate picture of the true number of offences committed, giving a more accurate view than official records and denotes it as his key hypothesis in this report, although he understands there are still some doubts about the validity of self-reports of offending due to obvious reasons such as concealing information or exaggerating or simply forgetting the details. In relation to our study we had no official records or arrest data to compare to, in order to discover how accurate our findings were but we discovered that self-reports provide a crucial and possibly highly accurate view into the offence of VPT and possible other cybercrimes. With this data we can get an insight into a species of criminal which has never been documented before, however we still understand the doubts that Farrington [25] states about offenders concealing of information or exaggerating certain answers however in our expert opinion we believe our data to be free from these limits with little or no outliers in the responses. This study provided much benefit to the research methodology we used for our study but it did not focus on self-reporting of cyber-criminals specifically, more over a focus on physical act and traditional crimes such as burglary.

## 3.7 Conclusion

One of the main advantages of the internet is the ability and ease in which users can access and share content electronically, but unfortunately it has become one of its weaknesses [4] creating a serious cyber security issue especially when it comes to VWEs and virtual property items. With the popularity of the internet, new ways have been conceived to conduct traditional crime in an electronic fashion, which has opened up a whole new wave of criminal activities such as VPT; the merging of burglary and VWEs.

This study gave results from a survey study on virtual property thieves. This survey examined the crime of VPT from an offenders point of view, our sample size for this study was one hundred (n=100) self-proclaimed offenders of VPT.

There were a number of limitations of this study which consisted of: any time when conducting survey studies that are anonymous in nature and respondents are of questionable character (the respondent group were self-proclaimed offenders of a cybercrime (however not recognized globally), results may be received which are not totally truthful or inconsistent; this is a factor in this study; however we benefited from very little 'unusual' or outlier responses which may indicate a truthfulness to the responses.

This survey study examined how offenders of VPT conduct the offence and related activities; in terms of virtual world profile analysis, thief profiling, theft practices, recovery and detection and security analysis. The results provide an insightful and concise view on how offenders conduct VPT as well as providing clarity to the VWE community of users as well as developers, publishers and operators. Users can view these results and view how offenders are breaking into accounts and conducting theft of virtual property items, and as a result become more aware with possibly attempting to secure themselves against any of these threats. Developers, publishers and operators can view these results and determine that this in indeed a problem for the VWEs they operate and look into how offenders are conducting the act of VPT and possibly spend time developing techniques to protect against the more popular methods used to conduct VPT as shown in Table I.

A surprising trend discovered from the results showed a clear cut profile of a virtual property thief. They appear to be mainly aged 20-24 years of age (n=37), living in the United States of America (n=35), they use VWEs for 5-7 hours a day (n=38) and have used VWEs for 2 years or more (n=42). The thieves appear to be young individuals, with a lot of free time on their hands, who have a lot of experience with VWEs.

When it came to analysing the sale of stolen virtual property items it was discovered from the results, offenders are stealing or have stolen for the most part 1000s of items over the lifetime of their activities (n=26) and unexpectedly these stolen items are being sold on primary markets (n=51) nearly as much as secondary markets (n=76), which shows that none of these markets are secure or have the ability to stop the sale of stolen items. This study discovered that the amount of real world money made from the sale of stolen virtual property showed that 23 offenders have made over 50,000 US dollars and the total offenders (n=100) in the study have accumulated over 5 million US dollars.

The results and analysis in this study takes a step towards proving that VPT offenders are conducting this offence at a substantial rate and making large amounts of real world money, however they appear to be lacking expert knowledge when it comes to conducting these hacking activities (ability to break into VWE accounts with n=63 low-medium self-rated ability and n=76 use the same computer and internet setup for each account they break into). The need to develop some kind of security measure or technique to stop these offenders is paramount in stopping this offence and providing a safe, secure and enjoyable environment for the VWE community.

## REFERENCES

[1]  N. Naik and P. Jenkins, "Discovering Hackers by Stealth: Predicting Fingerprinting Attacks on Honeypot Systems," in 2018 IEEE International Systems Engineering Symposium (ISSE), 2018, pp. 1-8.

[2]  P. Chiaramonte and J. Winter. (2011). Hacker Group Anonymous Threatens to Attack Stock Exchange. Available: https://www.foxnews.com/tech/hacker-group-anonymous-threatens-to-attack-stock-exchange

[3]  S. Cowley. (2012). FBI Director: Cybercrime will eclipse terrorism. Available: https://money.cnn.com/2012/03/02/technology/fbi_cybersecurity/

[4]  K.-K. Choo and R. Smith, "Criminal Exploitation of Online Systems by Organised Crime Groups," Asian Journal of Criminology, vol. 3, no. 1, pp. 37-59, 2008.

[5]  Linden.Research.Inc. (1999). Second Life Official Site. Available: http://lindenlab.com/

[6]  Y.-C. Chen, J.-J. Hwang, R. Song, G. Yee, and L. Korba, "Online Gaming Crime and Security Issue - Cases and Countermeasures from Taiwan," presented at the Proceedings of the International Conference on Information Technology: Coding and Computing (ITCC'05) - Volume I - Volume 01, 2005. Available: https://nrc-publications.canada.ca/eng/view/object/?id=a4a70b1a-332b-4161-bab5-e690de966a6b

[7]  A. Watters. (2010). Number of Virtual World Users Breaks 1 Billion, Roughly Half Under Age 15. Available: https://readwrite.com/2010/10/01/number_of_virtual_world_users_breaks_the_1_billion/

[8]  M. Eikenberry. (2011). Real Money Trade is a Billions Dollar a year Industry. Available: http://www.youtube.com/watch?v=rtZY3fVwlgw

[9]  E. Feldmann. (2008). Netherlands Teen Sentenced for Stealing Virtual Goods. Available: http://www.pcworld.com/article/152673/netherlands_teen_sentenced_for_stealing_virtual_goods.html

[10] D. McNeill. (2008). Virtual killer faces real jail after murder by mouse Available: http://www.independent.co.uk/life-style/gadgets-and-tech/news/virtual-killer-faces-real-jail-after-murder-by-mouse-972680.html

[11] J. Graham and B. Bowling, "Young People and Crime," in "Home Office Research Study 145," London: HMSO1995.

[12] A. E. Hernandez, "Self-reported contact sexual offenses by participants in the Federal Bureau of Prisons'sex offender treatment program: Implications for internet sex offenders," presented at the 19th Research and treatment conference of the association for the treatment of sexual abusers, San Diego, CA, 2000.

[13] R. W. Saputra, "A survey of cyber crime in Indonesia," in 2016 International Conference on ICT For Smart Society (ICISS), 2016, pp. 1-5.

[14]  P. N. V. Kumar, "Growing cyber crimes in India: A survey," in 2016 International Conference on Data Mining and Advanced Computing (SAPIENCE), 2016, pp. 246-251.

[15]  M. Rogers, N. D. Smoak, and J. Liu, "Self-reported Deviant Computer Behavior: A Big-5, Moral Choice, and Manipulative Exploitive Behavior Analysis," Deviant Behavior, vol. 27, no. 3, pp. 245-268, 2006/07/01 2006.

[16]  I. Deliu, C. Leichter, and K. Franke, "Collecting Cyber Threat Intelligence from Hacker Forums via a Two-Stage, Hybrid Process using Support Vector Machines and Latent Dirichlet Allocation," in 2018 IEEE International Conference on Big Data (Big Data), 2018, pp. 5008-5013.

[17]  R. Frank, M. Thomson, A. Mikhaylov, and A. J. Park, "Putting all eggs in a single basket: A cross-community analysis of 12 hacking forums," in 2018 IEEE International Conference on Intelligence and Security Informatics (ISI), 2018, pp. 136-141.

[18]  D. F. Criswell and M. L. Parchman, "Handheld Computer Use in U.S. Family Practice Residency Programs," Journal of the American Medical Informatics Association, vol. 9, no. 1, pp. 80-86, January 1, 2002 2002.

[19]  D. Rogers, G. Regehr, K. A Yeh, and T. Howdieshell, Computer-assisted Learning versus a Lecture and Feedback Seminar for Teaching a Basic Surgical Technical Skill. 1998, pp. 508-10.

[20]  J. Brand and S. Todhunter. (2015). Digital Australia Report 2016. Available: https://www.igea.net/wp-content/uploads/2015/07/Digital-Australia-2016-DA16-Final.pdf

[21]  Y. Kaneko, K. Yada, W. Ihara, and R. Odagiri, "How Game Users Consume Virtual Currency: The Relationship Between Consumed Quantity, Inventory, and Elapsed Time Since Last Consumption in the Mobile Game World," in 2018 IEEE International Conference on Data Mining Workshops (ICDMW), 2018, pp. 848-855.

[22]  D. Bates. (2010). Internet estate agent sells virtual nightclub on an asteroid in online game for £400,000. Available: http://www.dailymail.co.uk/sciencetech/article-1330552/Jon-Jacobs-sells-virtual-nightclub-Club-Neverdie-online-Entropia-game-400k.html

[23]  J. W. Mullins and R. Komisar, Getting to plan B: breaking through to a better business model. Harvard Business Press, 2009.

[24]  G. Tsakalidis, K. Vergidis, and M. Madas, "Cybercrime Offences: Identification, Classification and Adaptive Response," in 2018 5th International Conference on Control, Decision and Information Technologies (CoDIT), 2018, pp. 470-475.

[25]  D. P. Farrington, What Has Been Learned from Self-Reports About Criminal Careers And the Causes of Offending? London Home Office Online Report, 2001.

# Authors

**Dr Nicholas Patterson** is a Senior IT Lecturer at Deakin University with publications in Cyber Security, Blockchain and EdTech. He also runs his own consulting company, that brings advanced research into industry to increase capability and outcomes. His expertise is underpinned by a PhD which earned him the Alfred Deakin medal and a range of advanced micro-credentials in Critical Thinking, Problem Solving, Communication, Teamwork, Data Analytics, Innovation, Digital Literacy and Digital Learning.

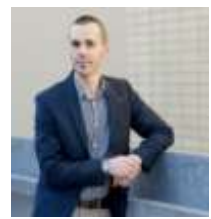

**Dr Michael Hobbs** is a Senior Lecturer at Deakin University, Australia. Dr. Hobbs was a researcher at Hewlett Packard labs from the years 2000 to 2002 and currently is the director of the Games Design and Development course at Deakin University. Dr. Hobbs research interests revolve around cloud computing, networking and security. He has been on the editorial board for many local and international conferences, as well as giving research presentations at conferences all around the world and has published more than 40 research articles in refereed international conferences and journals. Dr. Hobbs has been a successful mentor at Deakin University supervising many PhD students to completion from 2008 to 2013.

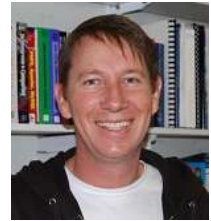

**Dr. Lei Pan** received his PhD in Computer Forensics from Deakin University, Australia, in 2008. He is currently a Senior Lecturer with the School of Information Technology, Deakin University. His research interests are cyber security and IoT. He has authored 50+ research papers in refereed international journals and conferences.

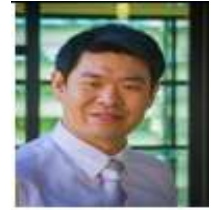

**Dr Frank Jiang** received his Ph.D. degree in the University of Technology Sydney in 2008. He also received the master's degree in computer science and gained the 3.5 years of postdoctoral research experience at the University of New South Wales (UNSW). His main research interests include Data-driven cyber security, Predictive analytics, biologically inspired learning mechanism and its application in the complex information security system, he has published over 90 highly reputed referred SCI/EI indexed journals and conferences articles.

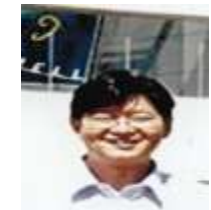